\documentclass[aps,prb,reprint,showpacs]{revtex4-1}  
\usepackage{dcolumn}
\usepackage[version=3]{mhchem} 
\usepackage{graphicx}
\usepackage{amsmath}
\usepackage{tikz}
\usepackage{mathtools}
\usepackage[utf8]{inputenc}

\begin{document}

\title{Itinerant magnetism in metallic \ce{CuFe2Ge2}}
\date{\today}
\author{K. V. Shanavas}
\email{kavungalvees@ornl.gov}
\author{David J. Singh}
\affiliation{Oak Ridge National Laboratory, 1 Bethel Valley Road, Oak Ridge, Tennessee 37831, USA}

\begin{abstract}
Theoretical calculations are performed to understand the electronic structure and magnetic properties of CuFe$_2$Ge$_2$. The band structure reveals large electron density $N(E_F)$ at the Fermi level suggesting a strong itinerant character of magnetism. The Fermi surface is dominated by two dimensional sheet like structures, with potentially strong nesting between them. The magnetic ground state appears to be ferromagnetic along $a$ and antiferromagnetic in other directions. These results show that CuFe$_2$Ge$_2$ is an antiferromagnetic metal, with similarities to the Fe-based superconductors; such as magnetism with substantial itinerant character and coupling between magnetic order and electrons at the Fermi energy.
\end{abstract}

\maketitle

\section*{Introduction}
Discovery of superconductivity in copper oxides~\cite{Fujita2012} and iron pnictides and chalcogenides~\cite{Stewart2011} has generated interest in the coexistence and interplay of magnetism and superconductivity.~\cite{Dai2012} The Fe-based materials in particular show a close association of superconductivity and antiferromagnetism, with at least partial itinerant character and coupling of magnetism to electrons at the Fermi surface.~\cite{Chen2008} 

Superconductivity in these systems is believed to be unconventional, in that it is mediated by antiferromagnetic spin-fluctuations.~\cite{Mazin2008} The large spin-fluctuations may arise as a consequence of nearness to a quantum critical point (QCP), which can also lead to non-Fermi liquid behavior, unusual transport and novel ground states.~\cite{Aguayo2004} There is evidence for this both from comparison of standard density functional calculations with experiments and spectroscopic probes.~\cite{Mazin2008a,Bondino2008}

In any case, it is of interest to look for other materials that share similar characteristics. In this manuscript we present theoretical investigations of electronic and magnetic properties of CuFe$_2$Ge$_2$, to serve as a precursor to future experimental studies. Our study is motivated in part by recent experimental results for YFe$_2$Ge$_2$ which indicate superconductivity along with highly enhanced Fermi liquid properties and scalings characteristic of a material near a magnetic QCP.~\cite{Zou2014} We identified CuFe$_2$Ge$_2$ as a compound with similar structural characteristics and, as discussed below, find many other similarities at the standard density functional level. CuFe$_2$Ge$_2$ is readily prepared by arc melting and crystallizes in orthorhombic structure with two formula units per cell.~\cite{Zavalii1987} Although it differs from the layered ThCr$_2$Si$_2$ structure of the Fe-based superconductors with 122 stoichiometry, bond lengths and interactions exhibit several similarities. Our calculated electronic band structure in the non-spin-polarized phase shows a large density of states at the Fermi level, consistent with itinerant character. Fermi surfaces in this system have a sheet like structure amenable to nesting and consequently to magnetic instabilities. Calculation of different magnetic configurations reveal significant variations in energies and $N(E_F)$ across them.

\section*{Methods}
The calculations reported in this manuscript are performed within density functional theory (DFT) using plane wave basis set and pseudopotentials with the Vienna {\it ab-initio} simulation package.~\cite{Kresse1993,Kresse1996} We use generalized gradient approximation (PBE-GGA) for exchange correlation.~\cite{Perdew1996} The calculations are carried out with the energy cutoff of 400 eV and {\bf k}-mesh of $16\times20\times12$ after carefully checking for convergence. We also verified some of our calculations with the all electron code Wien2k,~\cite{Blaha2001} to confirm that there are no artifacts of pseudopotential method.

\begin{figure}[ht]
{\includegraphics{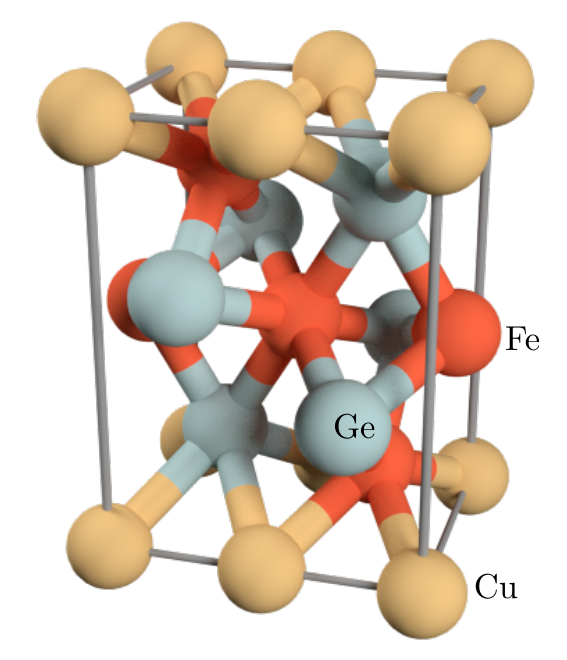}}
\caption{{\bf The crystal structure of CuFe$_2$Ge$_2$}. It shows the orthorhombic unitcell containing two formula units.}
\label{figstr}
\end{figure}

\begin{figure}[ht]
{\includegraphics{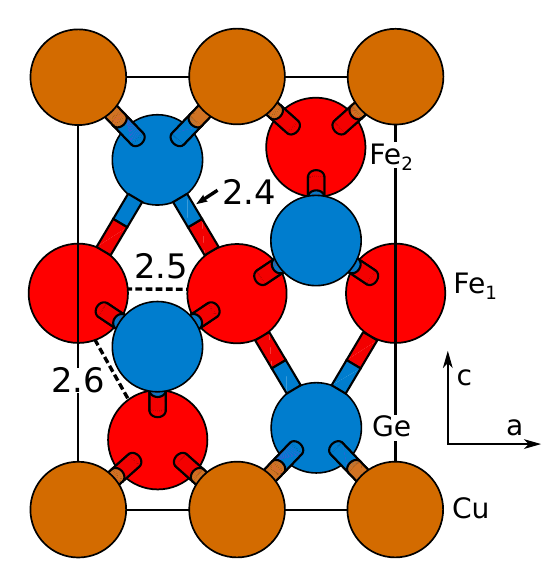}}
\caption{{\bf The projection of CuFe$_2$Ge$_2$ unitcell on to $ac$ plane.} Fe$_1$ (middle layer) are octahedrally coordinated. The atomic distances are marked in Angstrom.}
\label{figbond}
\end{figure}

The structure of CuFe$_2$Ge$_2$ was determined by Zavalii {\it et al} by x-ray powder diffraction on samples prepared through arc melting.~\cite{Zavalii1987} They found the material to crystallize in the orthorhombic structure, with space group $Pmma-D_{2h}^{16}$ and lattice constants (in~\AA) $a=4.98, b=3.97$ and $c=6.77$. As shown in Fig.~\ref{figstr}, the unitcell contains two formula units with Cu occupying the $2a$ sites. Fe occupies two inequivalent positions with the $2d$ ions (Fe$_1$) bonded octahedrally to surrounding Ge, while Fe at $2f$ sites (Fe$_2$) are bonded with four Cu ions on the $ab$ plane and two Ge on the $bc$ plane. Optimization of the experimental structure in the ferromagnetic phase yielded volume within 1\% and atomic coordinates within 0.1~\AA, which suggest that the theoretical equilibrium structure is very close to the experimental one. Note that nearest neighbor Fe-Fe distances are around 2.5~\AA\ (Fig.~\ref{figbond}), which is shorter than the 2.8~\AA\ in YFe$_2$Ge$_2$, and suggests that direct Fe-Fe interactions will be important in this system as well.

\section*{Results}
\subsection*{Electronic structure}
\begin{figure}[ht]
\scalebox{0.9}{\includegraphics{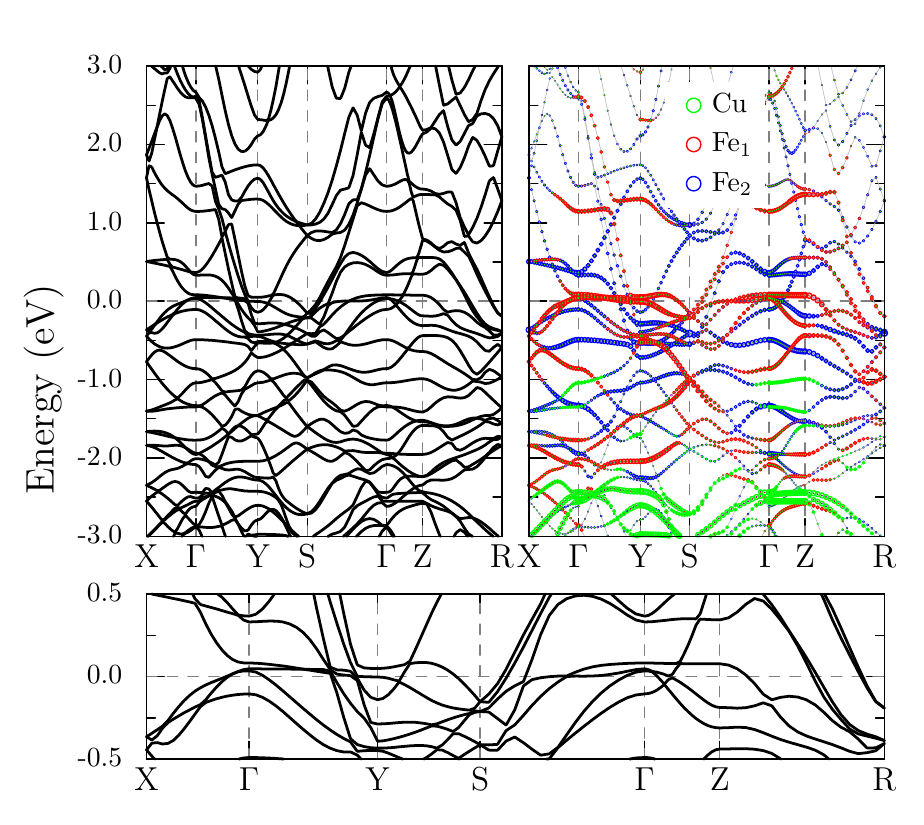}}
\caption{{\bf Electronic band structure of CuFe$_2$Ge$_2$ in the non-magnetic phase calculated using GGA.} The bottom panel zooms in on the states close to Fermi level which is set to 0 eV. The orbital character of the bands are shown by colored symbols in the top right panel. States close to Fermi level are dominated by Fe-$3d$ states.}
\label{figbs}
\end{figure}

\begin{figure}[ht]
\scalebox{1.0}{\includegraphics{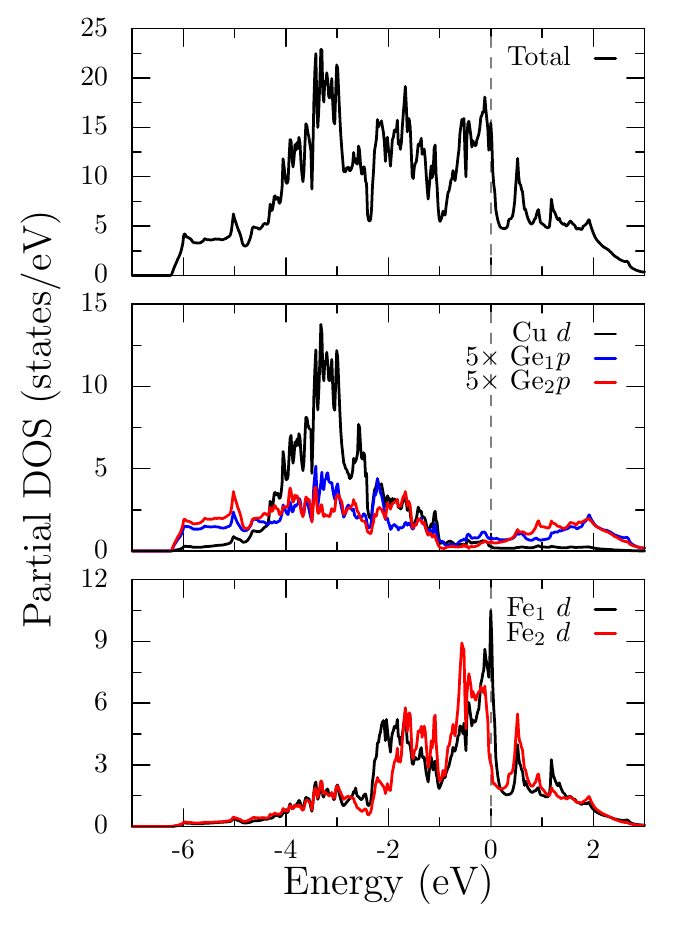}}
\caption{{\bf Total and partial density of states (DOS) for the non-magnetic calculation.} Cu-$d$ states are below the fermi level while Fe-$d$ and Ge-$p$ are partially occupied. Not shown are Fe and Cu $s$ states, that are empty.}
\label{figdos}
\end{figure}

As can be seen from the non-spin-polarized GGA bandstructure in Fig.~\ref{figbs}, the states close to Fermi level are dominated by Fe-$3d$ levels. Absence of strong crystal field splitting in the Fe-$d$ states point to the fact that the direct Fe-Fe interactions dominate, a character common to other iron-based superconductors.~\cite{Subedi2014,Singh2008} From the band character plot we can see that Fe$_1$ and Fe$_2$ exhibit different dispersions close to Fermi level. For the directions plotted, the Fe$_1$ bands remain relatively flat and this leads to a higher density of octahedral Fe$_1$-$3d$ levels near Fermi energy as can be seen from the partial density of states plotted in Fig.~\ref{figdos}. The Fe$_2$ bands show relative large dispersions along $\Gamma-Y$ and $\Gamma-Z$ directions, suggesting a three dimensional nature of the band structure. The Cu-$3d$ levels lie between -5 and -2 eV relative to Fermi level and thus are fully occupied, which can also be seen from Fig.~\ref{figdos}. Nearly all Fe and Cu $4s$ characters lie above the Fermi level. Counting the occupied states with different characters suggests nominal occupations of Cu $3d^{10}$, Fe $3d^{7.5}$ and Ge $4s^2 4p^3$.

\begin{figure}[h!]
\scalebox{1.0}{\includegraphics{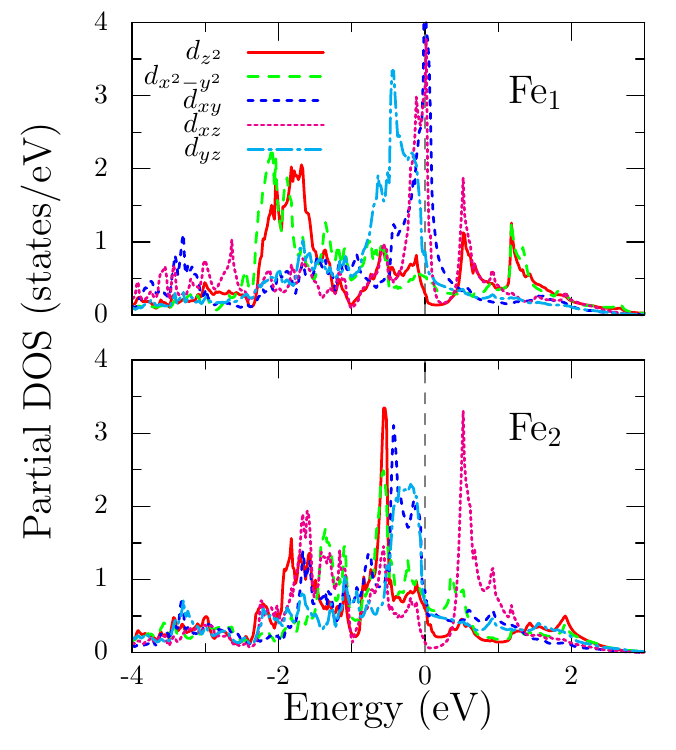}}
\caption{{\bf Partial density of states for Fe-$d$ states} Contributions from the $d$ states of Fe$_1$ and Fe$_2$ ions near the Fermi level from the non-magnetic calculation show that all states contribute significantly to the density of states. The crystallographic axes are taken as the coordinate system for calculation.}
\label{figfedos}
\end{figure}

The densities of states (Fig.~\ref{figdos}) also show features similar to YFe$_2$Ge$_2$ and the Fe-based superconductors~\cite{Singh2014,Subedi2014} such as the dip in the DOS just above $E_F$. The calculated electronic DOS at the Fermi level, $N(E_F)$, is also high at $N(E_F)=7.9$ eV$^{-1}$ per formula unit. This corresponds to a bare Sommerfeld specific heat coefficient $\gamma_{\rm bare}=18.6$ mJ/(mol K$^2$), which is in fact higher than the calculated value for YFe$_2$Ge$_2$. The contribution from different Fe sites to $N(E_F)$ are found to be significantly different; we get 5.2 eV$^{-1}$ and 1.6 eV$^{-1}$ respectively for Fe$_1$ and Fe$_2$. This suggest that the octahedrally coordinated Fe$_1$ has a higher tendency for itinerant magnetism in this material.  The contributions from different $d$ orbitals of Fe ions are shown in Fig.~\ref{figfedos}. As we can see, various states contribute towards density near the Fermi level similar to Fe pnictides.

\begin{figure}[ht]
\scalebox{0.27}{\includegraphics{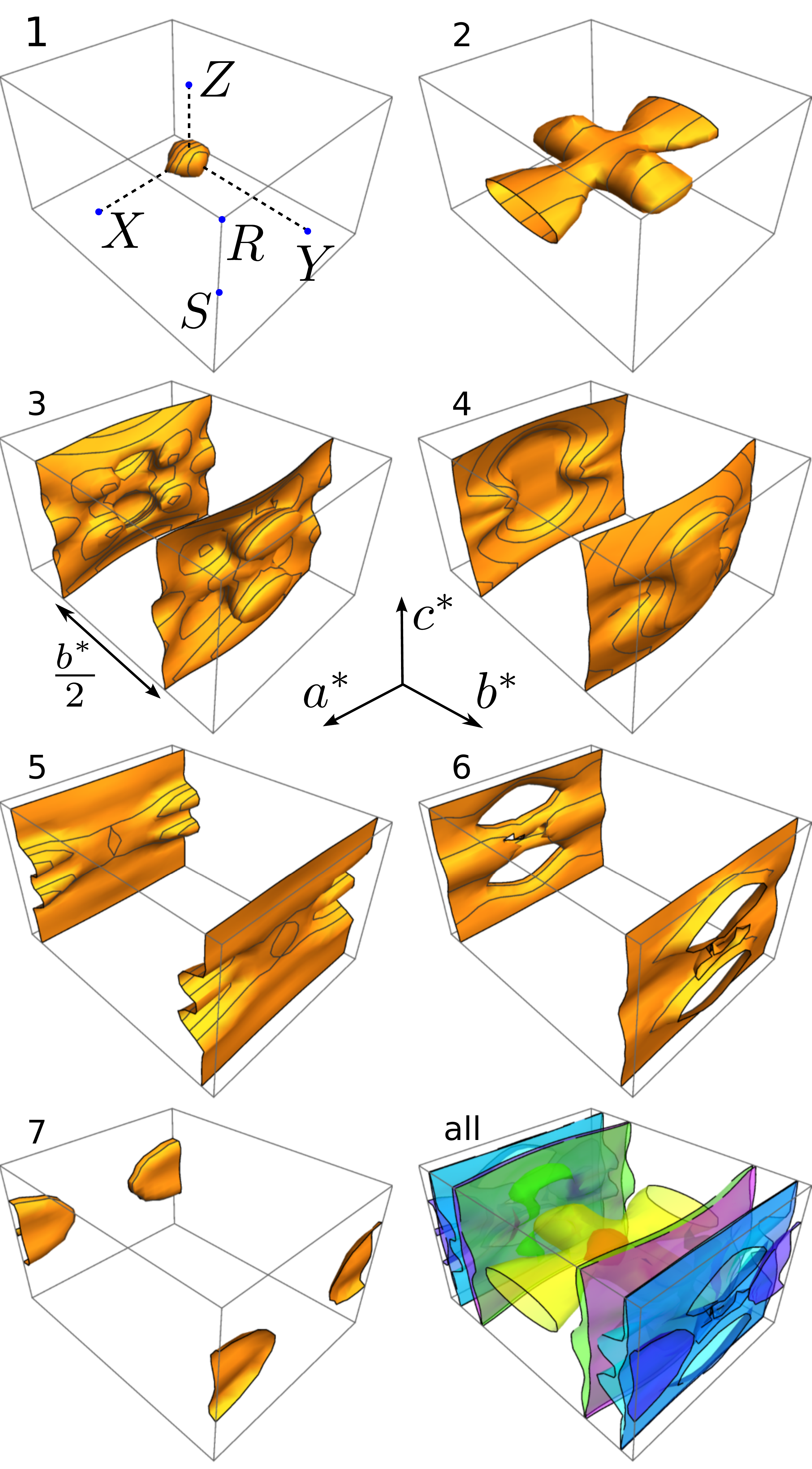}}
\caption{{\bf Calculated Fermi surfaces of CuFe$_2$Ge$_2$.} The high symmetry points are marked in the first panel. Seven bands with Fe-$d$ character cross Fermi level and bands numbered as 1,3,4,5 have strong Fe$_1$ character and bands 2,6,7 have strong Fe$_2$ character.}
\label{figfs}
\end{figure}

The calculated Fermi surfaces in the paramagnetic phase is shown in Fig.~\ref{figfs}. There are seven bands that cross Fermi level and all of them have strong Fe-$d$ character as can be seen from Fig.~\ref{figbs}. The Fermi surfaces are dominated by four mostly flat sheets in the $a^*b^*$ plane of the reciprocal space. Out of these the first three (3,4,5) have strong Fe$_1$ character. The closest of these, the pair marked by ``3'' in Fig.~\ref{figfs} are separated by half the reciprocal space distance $b^*$. We can expect strong nesting between these bands which may lead to antiferromagnetic coupling between Fe moments along the $b$ axis. In addition to the sheets perpendicular to the $b^*$ axis, there are also a hole pocket (1) at the $\Gamma$ point and an elongated disk shaped electron pocket (7) around $S$ and a dumbbell shaped structure (2) along $a^*$.

\subsection*{Magnetism}

The large density of states at the Fermi level $N(E_F)$, discussed earlier, implies that this system has magnetic instabilities as per the Stoner criterion.  Certain itinerant systems, such as YFe$_2$Ge$_2$, do not order magnetically even though DFT calculations predict magnetic ground state.~\cite{Singh2014,Subedi2014} Large spin-fluctuations are believed to be responsible for this effect, which happens in systems close to a QCP, and this leads to interesting new physics in these systems.~\cite{Zou2014} DFT is known to overestimate magnetic tendencies in itinerant systems close to QCPs because the exchange-correlation functionals commonly employed are based on the properties of the uniform electron gas which neglects the spin fluctuations associated with the QCP.~\cite{Aguayo2004} However, we are not aware of any magnetic measurements of CuFe$_2$Ge$_2$.

Thus, we carried out spin polarized calculations in several magnetic configurations and find that a ferromagnetic state is lower in energy by 102 meV/f.u. compared to non-spin-polarized state. A magnetic moment of about 1.4 $\mu_B$ develops on the Fe atoms and the $N(E_F)$ drops to about 4.1 eV$^{-1}$. For comparison, we also calculated other magnetic structures and the results are listed in Table.~\ref{taben}, and explained as follows. In the AF-$c1$, the Fe planes are arranged antiferromagnetically along the $c$ direction and ferromagnetic in other directions. In AF-$c2$ the Fe$_1$-Fe$_2$ coupling is AFM along $c$, but Fe$_1$-Fe$_1$ is FM. The AF-$b$ is AFM along the $b$ direction and we find that a similar AF-$a$ could not be stabilized. The AF-$C$ corresponds to checkerboard arrangement of moments in the $ab$ plane. Finally, AF-$G$ is obtained by starting from a different configuration and turns out to have the lowest energy. It is similar to AF-$c2$, except that it is also AFM along the $b$ direction.
 
\begin{table}[ht]
\caption{{\bf Magnetic properties from GGA.}}
\begin{tabular}{lcccc}
\hline\hline
Order & $m_{\rm Fe1}$ ($\mu_B$) & $m_{\rm Fe2}$ ($\mu_B$) & $E$ (eV/f.u.) & $N(E_F)$\\
\hline
NSP    & 0.00 &   0.00   &     0.00         &       7.91             \\
FM     & 1.44 &   1.40   &    -0.10         &       4.13              \\
AF-$c1$  & 1.28 &   1.22   &    -0.15         &       3.36    \\
AF-$c2$  & 1.32 &   1.30   &    -0.16         &       3.49    \\ 
AF-$b$   & 1.52 &   1.40   &    -0.10         &       4.72    \\ 
AF-$C$   & 1.45 &   1.41   &    -0.19         &       2.86    \\ 
AF-$G$   & 1.40 &   1.45   &    -0.21         &       3.04    \\ \hline
\end{tabular}
\begin{flushleft} For different magnetic ordering patterns studied (see text), the resulting magnetic moments $m$, energies and density at the Fermi surface $N(E_F)$ in units of eV$^{-1}$ per formula unit. The energy of paramagnetic case (NSP) is taken to be zero.
\end{flushleft}
\label{taben}
\end{table}

The large variations in energies and $N(E_F)$ across the ordered magnetic configurations suggest itinerant character of Fe moments. Note in particular that the difference in energy between different magnetic orders is as much as 0.11 eV/f.u., which is more than half the as much energy difference between a non-spin-polarized calculation and the lowest energy state and that the values of $N(E_F)$ also vary over a large range depending on the particular order. Moreover, the lowest energy configuration AF-$G$ is only 18 meV lower than the next lowest configuration. In our calculations, FM and AF-$b$ structures have similar energies, which suggest that magnetic exchange along $b$ direction is small. However, AF-$G$ is lower than AF-$c2$ by about 50 meV/f.u., while the difference between the two cases is AFM along $b$. This also suggests that a Heisenberg type model with nearest neighbor interactions will not fit this system.

Finally, we also calculated magnetic phases with local density approximation (LDA). It has been found that for itinerant systems near QCP, calculations with LDA predict much weaker magnetic tendencies.~\cite{Singh2014,Mazin2008a} This corresponds to a relative weakness of the momentum formation. Indeed, we find that the Fe moments in the FM configuration are 0.98 and 0.48 $\mu_B$ within LDA which are much smaller than the GGA moments of 1.44 and 1.40 $\mu_B$ respectively, as given in Table~\ref{taben}.

\section*{Discussion}

Our theoretical calculations on CuFe$_2$Ge$_2$ suggest that this material is an interesting itinerant magnetic system. The electronic structure shows similarities with those of the Fe-based superconductors. The Fermi surfaces show parallel sheet-like structures with potential nesting induced instabilities. Several magnetic configurations are stable, but the lowest energy is reached when Fe moments are ferromagnetic along $a$ and antiferromagnetic along $b$ and $c$ directions. On the scale of moment formation energy, there are significant variations in energies and densities of states at the Fermi level between different magnetic configurations, which also points to itinerant character in this system. We believe CuFe$_2$Ge$_2$ deserves further experimental investigation.

\section*{Acknowledgments}

This research was supported by the US Department of Energy, Basic Energy Sciences, Office of Science, Materials Sciences and Engineering Division.

\end{document}